\documentclass{eas}
\usepackage[dvips]{graphicx}

\newcommand{\Msun}{M$_\odot$}
\newcommand{\kms}{km~s$^{-1}$~}
\def\deg{\ifmmode^\circ\else$^\circ$\fi}  
\def\arcsec{\ifmmode {'' }\else $'' $\fi}  
\def\arcmin{\ifmmode {' }\else $' $\fi}    
\begin{document}

\title{The Accretion of Fuel at the Disk-Halo Interface} 
\author{M.E. Putman}\address{Dept. of Astronomy, Columbia University, 550 West 120th St, New York, NY 10027}
\author{J.E.G. Peek}\address{Dept. of Astronomy, University of California, Berkeley, CA 94720}
\author{F. Heitsch}\address{University of Michigan, Dept. of Astronomy, 500 Church St, Ann Arbor, MI 48109}
\begin{abstract}
We discuss the support for the cooling of gas directly at the disk-halo interface as a mechanism to continually fuel a galaxy.  This may be an important reservoir as there is not enough cold gas observed in galaxy halos and simulations indicate the existing cold clouds will be rapidly destroyed as they move through the surrounding halo medium.  We show possible evidence for a net infall of the WIM layer in the Milky Way, simulation results showing the recooling of warm clouds at the disk-halo interface, and GALFA HI data of small, cold HI clouds that could represent this recooling.   
\end{abstract}
\maketitle
\section{Introduction}
A typical spiral galaxy requires a source of ongoing incoming fuel to maintain its star formation and explain its stellar metallicity distribution (Chiappini et al. 2001; Sommer-Larsen et al. 2003).  The Milky Way is a specific example of a galaxy that will run out of star formation fuel within a few Gyrs without an ongoing source.  In addition, the fact that the amount of HI in the universe appears to remain flat since $z=3$, but the stellar content continues to increase, indicates the HI must somehow be continually replenished (e.g., Prochaska \& Wolfe 2009;  Putman et al. 2009; Hopkins et al. 2008).

Any external star formation fuel must enter the galaxy through its halo, and
be integrated into the disk through the disk-halo interface.
The primary candidates for the direct accretion of cold gas from the halo are the satellite galaxies and the clouds of hydrogen gas found around galaxies (e.g., Wakker \& van Woerden 1997; Thilker et al. 2004; Oosterloo, Fraternali \& Sancisi 2007).  In some cases the satellites and halo clouds may be directly related (e.g., Westmeier et al. 2008).   The problem with a spiral galaxy obtaining the majority of its star formation fuel via the direct accretion of cold halo gas, is neither the satellites or halo clouds appear to host enough fuel.   Accounting for completeness limits in the satellite counts of the Milky Way, dwarf satellites ($<10^{10}$ \Msun) have only brought in $\sim 0.2 M_\odot$ of fuel over the past few Gyrs (Grcevich \& Putman 2009).  Additionally, satellites are only capable of providing gas to the outskirts of galaxies, far from the star-forming regions of the disk where the gas is required (Peek 2009). The cold halo clouds that may or may not be related to the satellites also do not contain a large amount of mass and will provide $<0.5$ Gyr of future fuel unless somehow replenished (Putman 2006; Sancisi et al. 2008).

An additional source of star formation fuel is needed for galaxies beyond what can be obtained from cold halo clouds and satellites.  A possible reservoir is the extended warm-hot halo medium that surrounds our galaxy and the warm halo clouds within it.   This multi-phase halo medium reveals its presence around our Galaxy through pulsar dispersion measure studies, the structure of the HI halo clouds, and O~VI absorption and H$\alpha$ emission in the halo and at the disk-halo interface (Gaensler et al. 2008; Peek et al. 2007; Sembach et al. 2003; Putman et al. 2003; Haffner et al. 2003; Savage et al. 2003).  Some of this Galactic halo gas is likely to be the equivalent of the Mg~II and Ly$\alpha$ absorbers found in the vicinity of other galaxies (Gauthier et al. 2009; Kacprzak et al. 2008).   Though there is now direct evidence for this multi-phase medium around galaxies, how it could be subsequently utilized by a galaxy as star formation fuel has not been addressed.  In this paper we discuss how this process may reveal itself at the disk-halo interface.

\section{Accretion of Cold Gas?}

\begin{figure}
\begin{center}
\includegraphics[width=\textwidth]{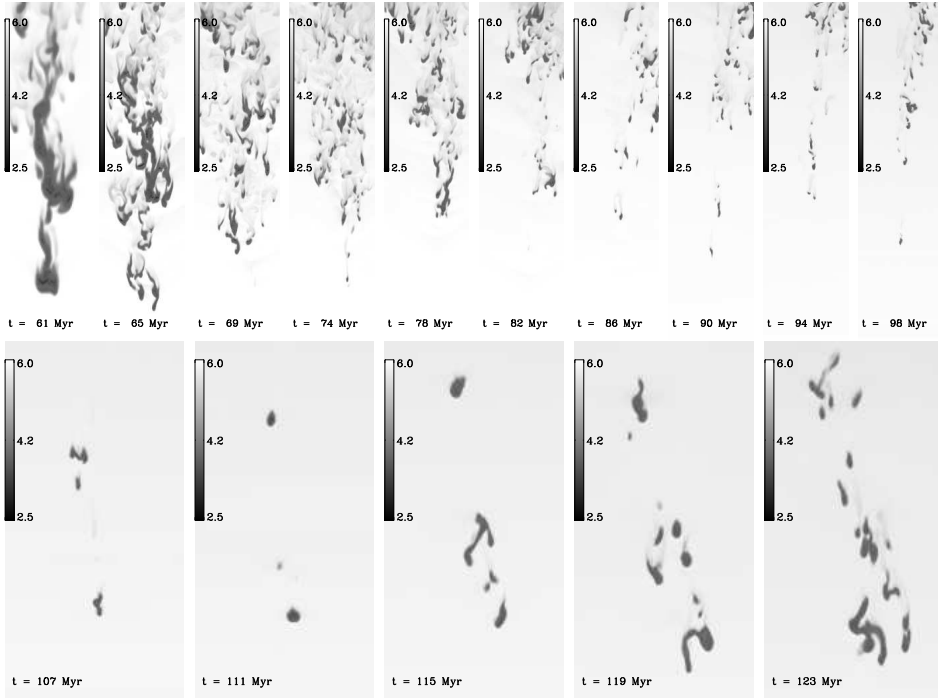}
\end{center}
\caption{\label{sim}{\em Top:} Time series of a disrupting HVC showing the final stages of the disruption (model Hc1b13a of Heitsch \& Putman 2009). The simulation is started at 0 Myr. The gray scale
                indicates the logarithm of the density-weighted temperature projected along the line of sight. The temperatures
                range between $10^2$ and $10^6$K. The box is 1 kpc wide. 
                Dynamical instabilities disrupt the cloud into small fragments.
                {\em Bottom:} Time series of recooling cloudlets in the same model starting at 107 Myr and including only the lower half of the top field.  
                Due to the increasing pressure as the cloud falls toward the Galactic plane, the fragments start to re-acquire gas
                and develop into a string of small cloudlets.}
\end{figure}

A limited amount of cold HI gas is observed in galaxy halos, and in addition much of this gas
may be unlikely to make it to the disk in cold form.  
The top series of panels in Figure~\ref{sim}
show the evolution of a typical halo cloud as it falls through an isothermal halo medium (Heitsch \& Putman 2009).  
These simulations demonstrate that for a range of cloud masses and for a typical halo medium, the
clouds lose 90\% of their cold ($< 10,000$ K) gas content after 
traveling less than 10 kpc through the halo.   Distance constraints have recently placed many HI halo clouds (or high-velocity clouds; HVCs) at $z-$heights in the range of $5-10$ kpc (Wakker 2001; Thom et al. 2008), or $>$ 60 kpc in the case of the Magellanic Stream.  This indicates many of the observed clouds in the Galactic halo will not make it to the disk in cold form (see also Bland-Hawthorn et al. 2007).  They may therefore merge into the gaseous thick disk as warm clouds or become part of the multi-phase extended diffuse halo.  

\begin{figure}
\begin{center}
\includegraphics[height=3in, bb=0 210 620 730,clip]{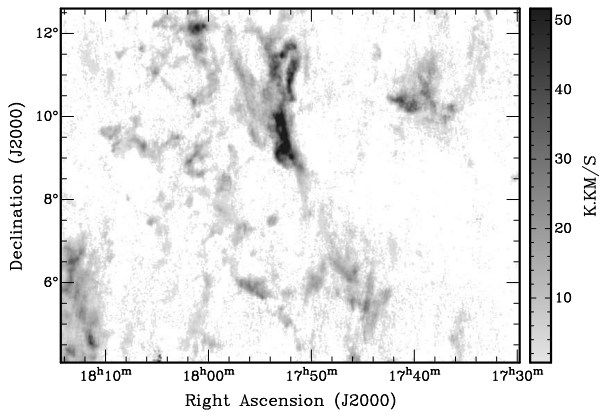}
\end{center}
\caption{GALFA HI data showing the tail of Complex C as a series of smaller clouds (Hsu et al. 2009).  This is an integrated intensity map including LSR velocities from -185 to -72 km s$^{-1}$.  The center of this image in Galactic coordinates is approximately $l,b =$ 18\deg, 32\deg.  The beam of Arecibo at 21-cm is 3.4\arcmin, which is $\sim$10 times smaller than the LDS data that is traditionally used to make maps of Complex C.\label{compc}}
\end{figure}

The larger the HI cloud, the further the cold gas is able to journey towards the disk, indicating that larger HVC complexes like Complex C (M$_{\rm HI} = 5 \times 10^6$~\Msun) will accrete largely in cold form.  Though this is partially the case, large complexes are often seen to break up into a series of smaller clouds when observed at higher resolution, especially at the tail of the complexes (see Figure~\ref{compc}, Putman et al. 2003, and Stanimirovic et al. 2008).   In any case, halo clouds with M$_{\rm HI}< 10^{4.5}$~\Msun, moving through a halo medium with densities consistent with observed constraints, will be largely devoid of cold gas by the time they reach the disk (Heitsch \& Putman 2009).  This encompasses the majority of the individual HVCs in the Galactic halo.

\section{Accretion of Warm Gas?}

Does the cooling of warm gas at the disk-halo interface serve as an important fueling mechanism?   This gas would presumably be a combination of the incoming warm clouds and feedback processes within the Galaxy.  It also may represent a gradual cooling of the warm-hot gas present in an extended Galactic halo.
Observationally we know there is a layer of warm gas that extends to $z-$heights of approximately 1-2 kpc though the detection of H$\alpha$ and other emission and absorption lines (Reynolds 1993; Gaensler et al. 2008; Savage \& Wakker 2009).  This layer of warm ionized medium (WIM) has been detected around our Galaxy and others (e.g. Rand 1996; Rossa \& Dettmar 2000;  Hoopes et al. 1999), with evidence for a slightly lagging rotational component with increasing $z-$height in those edge-on galaxies studied to date.   A layer of warm-hot gas is found to extend to higher scale heights of 3-5 kpc in our Galaxy through absorption line observations (e.g., O~VI, C~IV; Savage \& Wakker 2009; Sembach \& Savage 1992; Shull \& Slavin 1994).

\begin{figure}
\begin{center}
\includegraphics[height=3in, bb=0 400 720 1230,clip]{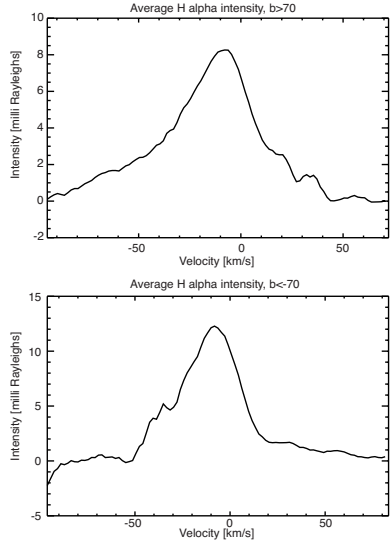}
\end{center}
\caption{The average velocity of all H$\alpha$ WHAM data at $b>70$\deg~(top) and $b<-70$\deg~(bottom) (Haffner et al. 2003).  The negative offset looking both above and below the plane is indicative of a net infall of the WIM layer.\label{wim}}
\end{figure}

A possible test of the WIM representing an infalling, cooling layer of fuel is to examine the kinematics of the WIM above and below us in the Galaxy.  Indeed, when examining the WIM at $|b|>70$\deg, it shows a net, low-velocity infall to both the north and south at $\sim$10 \kms (Haffner et al. 2003). This is found consistently in the WHAM survey, and shows the expected sin (b) dependence of an infalling layer (see Figure~\ref{wim}; Reynolds et al. 2004).
Though this may represent evidence for an infalling layer of warm gas and is consistent with an infall rate of $\sim$1 \Msun$/$yr, it is difficult to disentangle possible local effects from an overall, Galaxy-wide inflow, given our position within the Galaxy. The fact that the HI gas also shows a (less) negative offset towards both poles clouds the interpretation of an infalling WIM layer, although we note that while both show signs of inflow, variations in this inflow amplitude are not correlated along lines of sight and thus are unlikely to be connected physically. The HI gas most likely represents motion within more local gas, but needs to be better understood before making strong claims about an infalling WIM layer.

\subsection{Recooling of Warm Clouds}

\begin{figure}
\begin{center}
\includegraphics[height=3in, bb=0 180 620 730,clip]{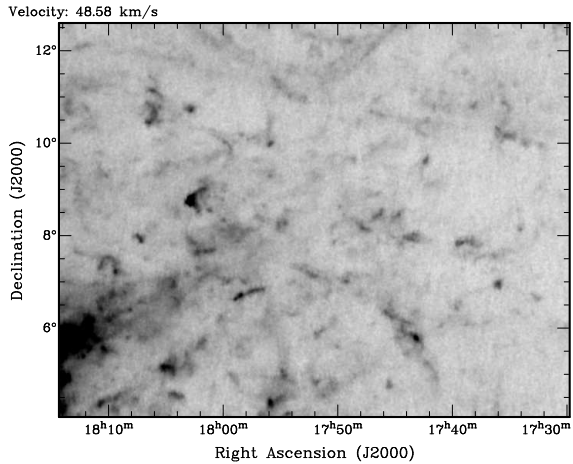}
\end{center}
\caption{Small, cold clouds at the disk-halo interface over a $\sim$3 \kms channel of GALFA data at V$_{\rm LSR} = 48.6~$\kms.   This is the same region as Figure~\ref{compc} centered at $l,b \sim$ 18\deg, 32\deg, but at a different velocity. \label{accrete}}
\end{figure}

The disruption of the HVCs represented in the top panels of Figure~\ref{sim} results in leftover warm fragments.   As these fragments move into a higher density surrounding medium, they become buoyant and begin to recool and grow (Heitsch \& Putman 2009).   This is shown in the bottom panels of Figure 1 by zooming in on the fragments at the bottom half of the top panels for later time steps.
These recooling fragments may be related to the discrete cold HI clouds being found at the disk-halo interface by numerous HI surveys (Lockman 2002; Ford et al. 2008; Stanimirovic et al. 2006).
These small clouds rotate with our Galaxy, but remain of unknown origin.  They most likely extend up to 3 kpc, and are therefore embedded in the WIM.  Figure~\ref{accrete} shows data from the Galactic Arecibo L-band Feed Array (GALFA) HI Survey at a velocity of 48.6~\kms showing a series of these clouds at the disk-halo interface.   Figure~\ref{accrete} is the same region as the tail of Complex C shown in Figure~\ref{compc}, but at a different velocity.  The accretion of gas from Complex C may play a role in the abundance of discrete cold clouds in this region.\footnote{Gas in this region was attributed to the Ophiuchus superbubble by Pidopryhora et al. (2007).} The accretion of the complex may include some warm clouds that are beginning to re-cool, or the complex may be applying additional pressure on the disk-halo interface causing denser concentrations to begin to cool.  Complex C, at this location and a constant distance of 10 kpc, has a $z-$height of only $2-3$ kpc and indeed would therefore be located at the disk-halo interface.  The velocities of Complex C also approach those of Galactic emission in this region (Hsu et al. 2009).

\section{Summary}

Neutral hydrogen surveys of cold gas around the Milky Way and other galaxies are not finding the quantity of fuel required to maintain a galaxy's star formation.   Combined with the flat $\Omega_{\rm HI}$ over cosmic time, this indicates the HI in galaxies must be continually replenished from other sources.  The gradual cooling of a halo medium onto a galaxy's disk would be a method of gas accretion that is not directly obvious and could potentially provide the needed star formation fuel for an extended period of time.   The cooling gas could be a combination of warm halo clouds, feedback material, and the gradual cooling of the extended warm-hot halo.   

The Milky Way is the only galaxy for which we can examine the disk-halo interface in detail.  Peering above and below us we see possible evidence for a net infall of the WIM layer (Figure~\ref{wim}) and discrete HI clouds that may represent the cooling of warm halo clouds (Figure~\ref{accrete}).  This is consistent
with results from simulations that indicate the cold gas from halo clouds (e.g., Figure~\ref{compc}) and small satellites is
unlikely to make it to the disk in cold form, and the remnant cloud fragments may recool at the disk-halo interface as they slow and encounter denser gas (Figure~\ref{sim}).  Cold gas may still reach the disk directly with the occasional merger with larger satellites (e.g., the Magellanic Clouds for the Milky Way and M33 for Andromeda (Putman et al. 2009)).  How much large satellites will immediately fuel the disk rather than first feeding the warm-hot halo will depend on 
their trajectories and the efficiency of the stripping process.


\end{document}